\documentclass{cjpsuppl}% for LaTeX 2e
\usepackage{graphicx}
\begin{document}
\title{Measuring the MSSM Lagrangean}
\authori{Tilman Plehn}
\addressi{CERN, Theory Division, Department of Physics, 1211 Geneva 23, Switzerland}
\authorii{}    \addressii{}
\authoriii{}   \addressiii{}
\authoriv{}    \addressiv{}
\authorv{}     \addressv{}
\authorvi{}    \addressvi{}
\headtitle{Measuring the MSSM Lagrangean}
\headauthor{Tilman Plehn}
\lastevenhead{Tilman Plehn: Measuring the MSSM Lagrangean}
%%%%%%%%%%%%%% Pro editory supplementu: %%%%%%%%%%%%%%%
\refnum{}%slouzi editorum pro evidenci; nakonec {}
\daterec{}
\suppl{A} \year{2004} \setcounter{page}{1}
%\firstpage{1}
%\lastpage{000}
%\makefirsttitle
%%%%%%%%%%%%%%%%%%%%%%%%%%%%%%%%%%%%%%%%%%%%%%
\maketitle

\begin{abstract}
  At the LHC, it will be possible to investigate scenarios for physics
  beyond the Standard Model in detail. We present precise total and
  differential cross section predictions, based on Prospino2.0 and
  SMadGraph. We also show to what degree the structure of the
  weak--scale supersymmetric Lagrangean can be studied, based on LHC
  data alone or based on combined LHC and linear collider. Using
  Sfitter we correctly take into account experimental and theoretical
  errors. We make the case that a proper treatment of the error
  propagation is crucial for any analysis of this kind.
\end{abstract}

\section{Introduction}

In the near future, we expect the LHC experiments to unravel the
mechanism of electroweak symmetry breaking and to search for new
physics at the TeV scale. Over many years it has been established that
for example supersymmetry can be discovered at the LHC, but also that
the supersymmetric partner masses can be measured in cascade
decays~\cite{Bachacou:1999zb}. Mass measurements at the percent
level, which can be supplemented with cross
section~\cite{Beenakker:1996ch} and branching fraction
measurements~\cite{Muhlleitner:2003vg} and with additional constraints
like the dark matter relic density~\cite{Polesello:2004qy}, will allow
us to determine weak--scale Lagrangean parameters. Mass measurements
at a future linear collider are typically at least one order of
magnitude better than their LHC counterparts. More importantly, the
sectors of the MSSM probed by the two machines nicely complement each
other. A proper combined analysis~\cite{Lafaye:2004cn} covers the
entire MSSM spectrum and probes the complete weak--scale
Lagrangean. It might even allow for an extrapolation to high
scales~\cite{Lafaye:2004cn,Blair:2002pg} where the structures of SUSY
breaking become visible.

\section{Next-to-leading order total cross sections}

\begin{figure}[t]
\begin{center}
\includegraphics[width=9.0cm]{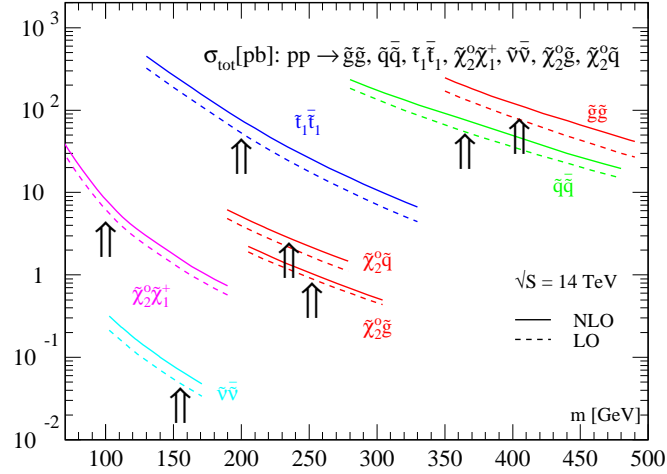} 
\end{center} \vspace*{-10mm}
\label{fig:propaganda}
\caption{Leading order and NLO production cross sections at the 
LHC as a function of the average final state mass. The arrows 
indicate a typical SUGRA scenario.}
\end{figure}

One basis for the inclusive search for supersymmetry at hadron
colliders is precise predictions of all production cross
sections. Similar to their Standard Model counterparts, SUSY-QCD cross
sections are plagued by large theoretical errors, which can for
example be observed in the renormalization and factorization scale
variations. The Tevatron searches use next-to-leading order cross
section predictions --- until now unfortunately only ruling out parts
of the SUSY parameter space. The large set of NLO cross sections at
hadron colliders shown in Fig.~\ref{fig:propaganda} can be computed
using the publicly available computer program
Prospino2.0~\cite{Beenakker:1996ch}. In addition to squark and gluino
production, pair production of neutralinos and charginos, which can
decay into trilepton final states, has been included.  Recently, we
have seen that in split-supersymmetry models these Drell-Yan type
processes together with the production of long-lived gluinos are the
main SUSY signals at the LHC~\cite{Kilian:2004uj}.  This feature is
not limited to the extreme case of split supersymmetry. Cascade decays
of squarks can already be decoupled for only slightly enhanced squark
masses. A small hierarchy between gauginos and scalars appears for
example in gravity mediated models with anomaly mediated gaugino
masses, where this hierarchy alleviates the problems in the SUSY
flavor sector~\cite{Wells:2003tf}.

\begin{figure}[t]
\begin{center}
\includegraphics[width=7.0cm]{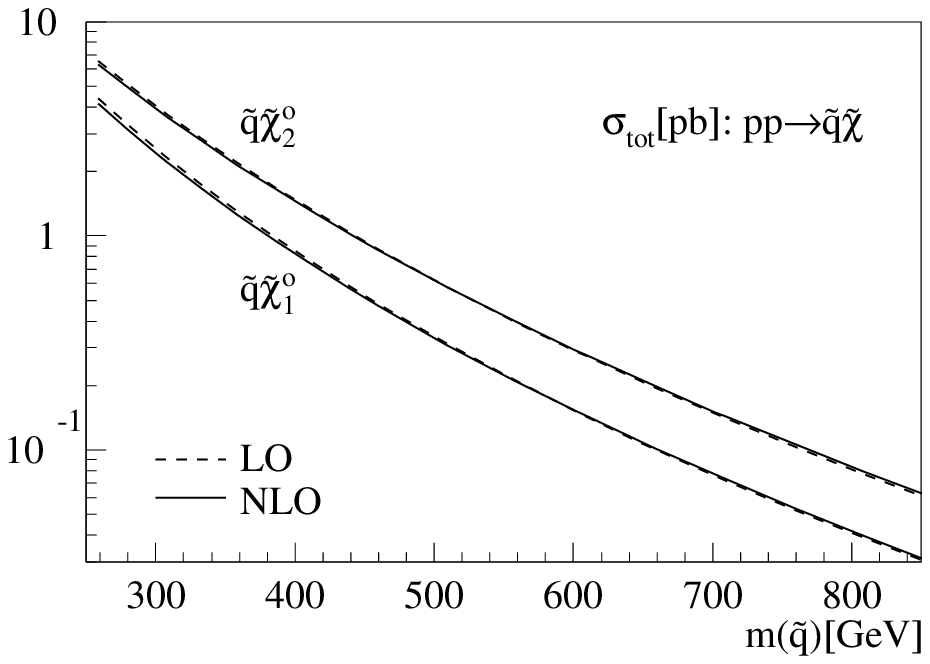} \\
\includegraphics[width=7.0cm]{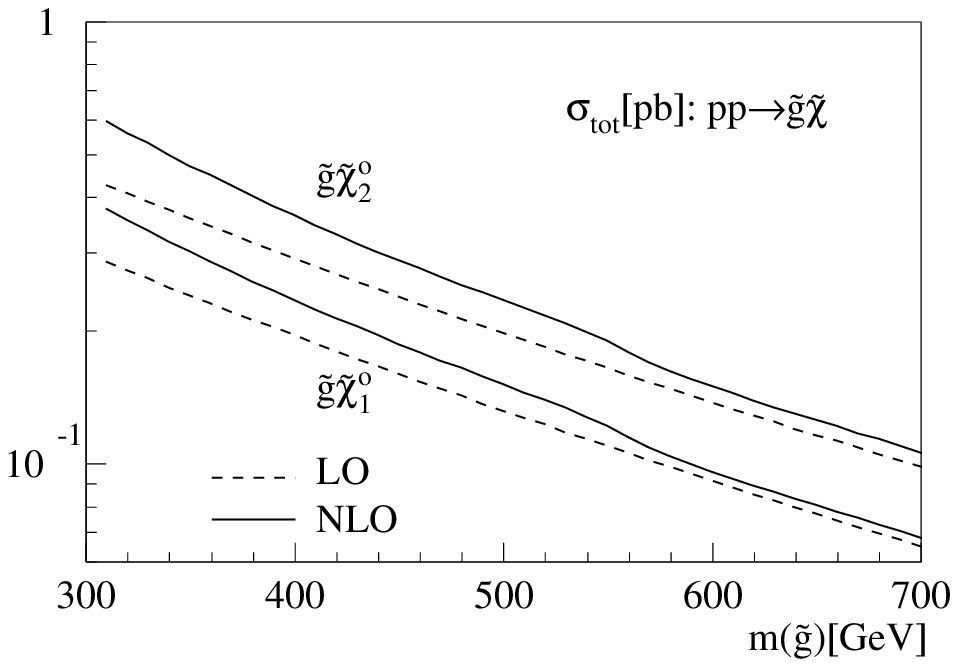} 
\end{center} \vspace*{-10mm}
\label{fig:asso}
\caption{Leading order and NLO production cross sections for the
associated production of neutralinos with a gluino or a squark.  The
MSSM parameters are fixed to the SPS1a data point, and the squark mass
or gluinos mass are varied independently in the graphs.  The central
parameters are $m_{\tilde{q}}=559$~GeV and
$m_{\tilde{g}}=609$~GeV. The two lightest neutralino masses are 96 and
179~GeV.}
\end{figure}

In this paper we present new results for the associated production of
charginos and neutralinos with
gluinos~\cite{Spira:2002rd,Berger:2000iu} and
squarks~\cite{my_talk}. After including these two classes of
production processes, the list of cross sections available in
Prospino2.0 is complete. The same way as for all processes shown in
Fig.~\ref{fig:propaganda} we compute the complete SUSY-QCD corrections
to the leading order processes $pp \to \tilde{q} \tilde{\chi}$ and $pp
\to \tilde{g} \tilde{\chi}$. The results are available for the
Tevatron and for the LHC. A technical complication is the correct
subtraction of intermediate particles: the NLO contribution $pp \to
\tilde{q}\tilde{\chi}$ includes an intermediate gluino with a
subsequent decay $pp \to \tilde{g} \tilde{\chi} \to (q\tilde{q})
\tilde{\chi}$, plus intermediate squark pair production $pp \to
\tilde{q}^* \tilde{q} \to (q\tilde{\chi}) \tilde{q}$. Of course, we
could regularize this on-shell divergence using a Breit--Wigner
propagator, but this would lead to double counting between
$\tilde{q}\tilde{\chi}$ production, $\tilde{g} \tilde{\chi}$
production, and $\tilde{q}^*\tilde{q}$ production at the NLO level. To
avoid any double counting in the combined inclusive SUSY samples, we
instead subtract the on-shell squark contribution in the narrow width
approximation~\cite{Beenakker:1996ch}. This procedure is uniquely
defined and allows us to naively add the different processes without
having to worry about double counting at all, when including NLO
effects to all $(2 \to 2)$ production processes\footnote{We use the
same procedure when combining charged Higgs production $pp \to tH^-$
and top pair production $pp \to t\bar{t} \to t (\bar{b}H^-$ at the
LHC~\cite{Berger:2003sm}.}. In the top panel of Fig.~\ref{fig:asso} we
see that this prescription leads to a smooth definition of the NLO
cross section for the process $pp \to \tilde{q}\tilde{\chi}$ around
the threshold $m_{\tilde{q}} = m_{\tilde{g}}$.  Exactly the same way
we subtract on-shell contributions $\tilde{g} \tilde{q}$ and
$\tilde{q} \tilde{\chi}$ from the NLO contributions to $pp \to
\tilde{g}\tilde{\chi}$, shown in the lower panel of
Fig.~\ref{fig:asso}. A small effect in the NLO cross section survives:
above the threshold $m_{\tilde{g}} = m_{\tilde{q}}$ the NLO
corrections decrease, because here the possibly on-shell intermediate
state is actually on-shell and is therefore subtracted. Below
threshold, these channels are still off-shell, but give a sizeable
contribution to the cross section. Owed to the smaller collider
energies, these thresholds are even smoother at the Tevatron.

\section{Hard jet radiation in association with SUSY-QCD}

\begin{figure}[t]
\begin{center}
\includegraphics[width=7.0cm]{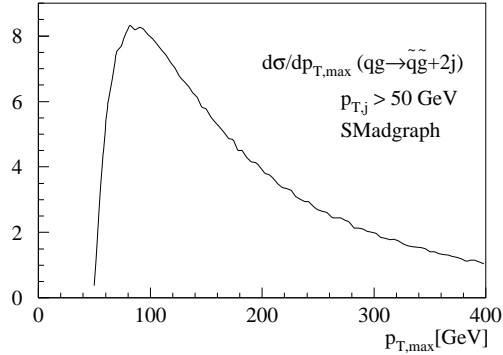} 
\end{center} \vspace*{-10mm}
\label{fig:smadgraph}
\caption{The transverse momentum spectrum of the harder jet for squark
and gluino production at the LHC, in association with two hard jets.
The phase space for the jets is limited to $p_T>50$~GeV.}
\end{figure}

A large fraction of the weakly interacting SUSY partner spectrum
appears as intermediate states in cascade decays of squarks and
gluinos at the LHC. Because of the possibly large number of squarks
and gluinos produced, the masses of these intermediate particles can
be reconstructed from measured edges and thresholds in the decay
cascades~\cite{Bachacou:1999zb}. However, most of these edges and
thresholds involve jets, and it is crucial to know which of the jets
in the event come from: (i) the correct position in the cascade, (ii)
another decay step in the cascade, or (iii) additional jet radiation
in the event. As long as the mass differences between gluino, squarks
and neutralinos are large, the cascade jets are required to be very
hard. Simulating the additional jet with Pythia and Herwig implicitely
assumes that all additional jets are fairly soft --- therefore it is
not surprising that they do not enter the analysis as candidates for
cascade jets after the appropriate $p_T$ cuts.

A typical SUSY analysis might require three or more hard jets with a
staggered $p_T$ cut of 150, 100, 50~GeV. We use
SMadGraph~\cite{smadgraph} to compute the rate and the $p_T$
distributions for additional hard jets from the matrix element $pp \to
\tilde{q} \tilde{g}+X$. We focus on two additional hard jets and limit
their phase space to $p_T>50$~GeV, to allow them to enter the SUSY
cascade analysis. Softer jets will increase the jet activity but will
not appear as a candidate for a cascade jet. The suppression of the
rate per additional hard jet of this kind is of the order of
1/2...2/3, so two additional jets reduce the total cross section for
$\tilde{q} \tilde{g}$ to around 1/3 of the leading order cross
section $pp \to \tilde{q} \tilde{g}$. This large factor is not
unexpected, even though the NLO $K$ factor for this process at the LHC
is well under control~\cite{Beenakker:1996ch}.

In Fig.~\ref{fig:smadgraph} we show the $p_T$ spectrum of the harder
of these two jets. It peaks around 100~GeV and decreases only very
slowly. This means that we cannot apply a $p_T$ cut on jets to remove
all jet radiation from the event sample. Instead, we will have to find
ways to deal with this new combinatorical error from additional matrix
element jets. While this jet radiation is not likely to become a
problem for the discovery of SUSY or the discovery of cascade decays,
it will soften the edges and thresholds and make it harder to fit the
masses which appear inside the cascade. In particular at the LHC, all
systematical and theoretical errors have to be taken into account to
quantify the precision with which we can measure the SUSY particle
masses and extract the weak-scale parameters in the MSSM Lagrangean
--- which we will discuss in the next section.

\section{Extracting the MSSM Lagrangean}

From a theorist's point of view the measurement of masses, cross
sections and branching fractions is only secondary. The relevant
parameters we need to study the structure of weak-scale supersymmetry
and possibly learn something about SUSY breaking are the parameters in
the Lagrangean. The neutralino and chargino sector is a
straightforward example: the two mass matrices are determined by the
weak gaugino mass parameters $M_{1,2}$, the Higgsino mass parameter
$\mu$, and by $\tan\beta$ (plus the Standard Model gauge
sector). These four parameters can be extracted out of six mass
measurements, but some of the mass measurements can also be traded for
total cross section or asymmetry measurements at a future linear
collider. Masses at a linear collider are the most precisely measured
observables.  Cross section measurements at the LHC are typically much
weaker even than mass measurements at the LHC because of the large
theoretical and systematical errors and are therefore not likely to be
helpful for the parameter extraction..\smallskip

As a simple starting point we assume gravity mediated supersymmetry
breaking and fit the four model parameters listed in
Tab.~\ref{tab:sfitter_sugra} to different sets of mass measurements.
We use SuSpect~\cite{Djouadi:2002ze} to compute the weak-scale MSSM
parameters. The experimental errors on the mass
measurements~\cite{georg_report} represent cascade reconstruction at
the LHC~\cite{Bachacou:1999zb} and threshold scans at a linear
collider~\cite{Aguilar-Saavedra:2001rg}). The resulting absolute error
on the high-scale model parameters is $\Delta^{(1)}$. Even though we
assume a smeared set of measurements and the starting point of the fit
is far away from the SPS1a scenario, the best fit values always agree
with the `true' values. Therefore, we omit the central values in the
table. The theoretical error on the prediction of weak-scale masses
from high-scale SUSY breaking can be estimated comparing different
implementations of the renormalization group
evolution~\cite{Allanach:2003jw} or comparing the two-loop results
which we use with three-loop beta function running~\cite{Jack:2004ch}.
We use a relative error of $1\%$ for weakly interacting particles and
$3\%$ for strongly interacting particles. The situation is more
involved for the light Higgs mass, which can be computed using
weak-scale input parameters. Here, the theoretical error consists of
errors in the input parameters, dominantly the top mass uncertainty,
and of unknown higher order effects. We assume a conservative error of
3~GeV altogether. When we add the theory errors and the experimental
errors in squares they yield an error $\Delta^{(2)}$ on the extracted
model parameters. We have checked that including theoretical errors
not as Gaussian but without any preference for the known value, {\it
i.e.}  using a box shape, does not have a significant impact on our
results.

As a cross check for the theoretical errors $\Delta^{(2)}$ we start
with a mass spectrum from SuSpect, but fit the parameters using the
SoftSusy renormalization group evolution~\cite{Allanach:2001kg}. The
results are also shown in Tab.~\ref{tab:sfitter_sugra}. We see that
the central values of $m_0, m_{1/2}$ and $\tan\beta$ lie within the
error bands $\Delta^{(2)}$ obtained from the consistent SuSpect
fit. Moreover, the errors of the consistent fit and the error on this
combination of different implementations agree well. The central
values for $A_0$ are in agreement with the error for the SuSpect
fit, but they are sufficiently far away from the central value to
illustrate that we cannot really determine $A_0$ from the set of
masses, given our assumed theoretical errors.\smallskip

If we step back and have a look at the parts of the spectrum we can
observe at the LHC and at a linear collider we see that the linear
collider misses the gluino and most of the squarks, while the LHC
leaves wide open spaces in the neutralino and chargino
spectrum. Considering gaugino mass unification as one of the most
important predictions of gravity mediated SUSY we could say that a fit
to the SUGRA parameters for either the LHC alone or a linear collider
alone assumes SUGRA to then get out SUGRA. In contrast, combining the
two sets of collider data nicely covers the entire MSSM spectrum and
does allow us to actually test the SUGRA hypothesis.

However, it would be clearly preferable if we were able to measure
weak-scale MSSM parameters, first confirm that some new physics looks
like supersymmetry, and then extrapolate this established parameter
set to some SUSY breaking scale. Even though there should be enough
measurements, the extraction of the weak-scale Lagrangean is
technically challenging: fitting a complex parameter space can become
sensitive on the starting values, because of domain walls which the
fit procedure cannot cross. It is not guaranteed that the best fit is
always the global minimum. A method which is more likely to find the
global minimum is a grid over the entire parameter space.  However, if
the parameter space is too big the resolution might not be sufficient
to make sure the result is the best local minimum.  Accounting for
these respective shortcomings, Sfitter uses a combination of $\chi^2$
minimizations based on a grid and on a fit.  In a subspace of the MSSM
parameter space we use a grid and compute an appropriate subset of
measurements (for example the neutralino--chargino sector).  Starting
from this minimum we fit first the remaining parameters to all
measurements, and then the entire parameter set to make sure we find
the correct local minimum. Recently proposed approaches like genetic
algorithms or adaptive scanning methods~\cite{Allanach:2004my} should
allow us to further improve the coverage of the MSSM parameter
space.\smallskip

\begin{table}[t]
\begin{footnotesize}
\begin{tabular}{|l|rrr|rrr|rrr|}
\hline
SPS1a
&   \multicolumn{3}{c|}{LHC} &
    \multicolumn{3}{c|}{LC} &
    \multicolumn{3}{c|}{LHC+LC} \\

      &         $\Delta^{(1)}$    & 
                $\Delta^{(2)}$  & 
                SoftSusy             & 
                $\Delta^{(1)}$    & 
                $\Delta^{(2)}$  & 
                SoftSusy             & 
                $\Delta^{(1)}$    & 
                $\Delta^{(2)}$  &
                SoftSusy             \\
\hline
$m_0$      & 4.0  &  4.6  & 98  $\pm$ 4.6 & 0.09 & 0.70 & 99  $\pm$ 0.7 & 0.08 & 0.68 &  99 $\pm$0.8\\
$m_{1/2}$  & 1.8  &  2.8  & 253 $\pm$ 2.9 & 0.13 & 0.72 & 251 $\pm$ 0.7 & 0.11 & 0.67 & 251 $\pm$0.8\\
$\tan\beta$& 1.3  &  3.4  & 11.6$\pm$ 3.4 & 0.14 & 0.49 & 10.1$\pm$ 0.5 & 0.14 & 0.49 & 10.1$\pm$0.6\\
$A_0$      & 31.8 & 50.5  & 14.7$\pm$ 14.2& 4.43 & 13.9 & -45 $\pm$ 16  & 4.23 & 13.1 & -43 $\pm$ 17\\
\hline
\end{tabular}
\end{footnotesize} \vspace*{-2mm}
\label{tab:sfitter_sugra}
\caption{Fit results from Sfitter. For the consistent running the
  central values of the fit are omitted, because they match the SPS1a
  input within the given errors. The input values are 100~GeV,
  250~GeV, 10 and -100~GeV for $m_0, m_{1/2}, \tan\beta$ and
  $A_0$. The experimental error alone leads to the quoted absolute
  error $\Delta^{(1)}$. After including a theoretical error on the
  masses, the error on the fitted values increases to
  $\Delta^{(2)}$. The last column assumes a spectrum created with
  SuSpect and a fit based on SoftSusy.}
\end{table}

Results for the fit of the weak-scale MSSM Lagrangean we presented in
Refs.~\cite{Lafaye:2004cn,remi_talk}. Details on the corresponding
Fittino parameter determination are also
available~\cite{fittino_thesis}. The main features for the SPS1a
parameter point are: all three gaugino masses are covered by a
combined LHC and linear collider analysis. The squark sector is
determined by the LHC, a linear collider will likely be limited to the
light stop. The slepton masses are well measured at the linear
collider, while the LHC only sees a subset in the lepton invariant
mass edge in cascade decays. The (heavy) Higgsinos are challenging for
the linear collider and for the LHC, but their masses could be
replaced by cross section measurements at the linear collider. The
determination of the three heavy-flavor trilinear coupling parameters
poses a serious problem: $A_\tau$ can be measured at a linear collider
through a combination of stau mass and cross section measurements.
$A_b$ might not be measured at either of the colliders, and for $A_t$
it is crucial to control the theoretical uncertainties of the light
Higgs mass and the correlations with $\tan\beta$ and $\mu$. At the
moment, the most important missing pieces are the stop sector at the
LHC~\cite{Hisano:2003qu} and a direct measurement of $\tan\beta$ at
either collider. For large values of $\tan\beta$ a measurement of
$\tan\beta$ might be possible in the production of heavy Higgs bosons
at the LHC, where the production rate is proportional to
$\tan^2\beta$~\cite{Allanach:2004ub}. These questions will be
investigated when we extend our study beyond the low $\tan\beta$
parameter point SPS1a, at which point we will have to combine the
SUSY-Higgs sector with the SUSY particle production.

\section{Outlook}

As part of the preparation for LHC data and its interpretation, there
has been considerable progress in the collider phenomenology of new
physics in general and of supersymmetry in particular. Beyond the
discovery of supersymmetry we will for example be able to measure
particle masses from the kinematics of cascade decays at the
LHC. These measurements (combined with cross section and branching
fraction measurements) can be used to determine a large fraction of
weak-scale Lagrangean parameters. For all these measurements it is
indispensable to reliably predict total and differential production
cross sections. The proper tools Prospino2.0 and SMadGraph are (or
will soon be) publicly available.

Moreover, we might well be able to use weak-scale measurements to
extrapolate Lagrangean parameters to higher scales and probe
supersymmetry breaking patterns. For this purpose, additional
precision measurements from a future linear collider are particularly
useful. It is absolutely crucial that we treat theoretical and
experimental errors correctly in these renormalization group
analyses. Using Sfitter (and Fittino) we show the promise of these
approaches and also their limitation for example in the case of LHC
data without any linear collider measurements.\bigskip

{\small I would like to thank David Rainwater, Tim Stelzer, Dirk
  Zerwas, Remi Lafaye and my coauthors on the Prospino publications.
  Without their collaboration the results presented in this article
  would simply not exist. Moreover, I am grateful to the DESY theory
  group for their kind hospitality while I wrote this contribution.}

\end{document}